\makeatletter
\declare@file@substitution{revtex4-1.cls}{revtex4-2.cls}
\makeatother

\documentclass[notitlepage,nofootinbib,preprintnumbers,amssymb,superscriptaddress]{revtex4-1}

\usepackage{amsfonts,amssymb,amsmath,graphicx,color,bm}
\usepackage[linktocpage=true]{hyperref}
\usepackage{enumitem}
\usepackage{orcidlink}
\usepackage{multirow}
\usepackage{subcaption}
\usepackage{booktabs}
\usepackage{caption}
\usepackage{physics}
\usepackage{array}
\usepackage{float}
\usepackage[12pt]{extsizes}
\usepackage{hyperref}
\usepackage{url}

\hypersetup{
    colorlinks=true,
    citecolor=red,
    linkcolor=blue,
    urlcolor=blue,
}

\begin{document}

\title{Inflationary models with a quadratic relationship between the parameters of cosmological perturbations}

\author{Igor V. Fomin\orcidlink{0000-0003-1527-914X}}
\email{fomin\_iv@bmstu.ru}
\affiliation{Bauman Moscow State Technical University, Moscow, Russia}

\author{Vladimir L. Glushkov\orcidlink{0009-0004-8816-8103}}
\email{glushkov-vl@bmstu.ru}
\affiliation{Bauman Moscow State Technical University, Moscow, Russia}

\author{Evgenii S. Dentsel\orcidlink{0009-0000-0629-5004}}
\email{edentsel@bmstu.ru}
\affiliation{Bauman Moscow State Technical University, Moscow, Russia}

\author{Gevorg D. Manucharyan\orcidlink{0000-0003-0464-0040}}
\email{gdmanucharyan@bmstu.ru}
\affiliation{Sternberg Astronomical Institute, Moscow, Russia}

\author{Vyacheslav A. Sizov\orcidlink{0009-0003-0415-5587}}
\email{vasizov@bmstu.ru}
\affiliation{Bauman Moscow State Technical University, Moscow, Russia}

\date{\today}

\begin{abstract}
An approach to construct cosmological inflation models on the basis of a certain dependence of the scalar field evolution on the e-folds number is considered. The reconstruction of the model background parameters according to the kind of specific connection between the parameters of cosmological perturbations is proposed. By the slow-roll approximation, as functions of the e-fold number, the scalar field potential and the Hubble parameter are determined. The logarithmic dependence between the scalar field evolution and the number of e-folds is investigated as the example. The derived model was found to obey a quadratic law between the parameters of cosmological perturbations. On the basis of suggested approach, a new cosmological inflation model with an exponential potential is proposed. This complies with existing observational constraints for both background and perturbation spectrum parameters. A comparison of obtained results with the solutions obtained by the symbol regression is made. The restrictions for arbitrary parameters in the model with logarithmic potential are also considered. The proposed form of the scalar field potential is consistent with one of the optimal potentials obtained when using machine learning methods to analyze the correspondence of inflationary models to observational data and gravitational waves contribution to the anisotropy and polarization of the CMB.
\end{abstract}


\maketitle

\section{Introduction}
The Big Bang theory faces several issues, which can be resolved by assuming the existence of an inflationary stage. For example, it allows to take into an account the formation of Universe large-scale structure. A crucial assumption in solving such problems is the existence of an inflationary stage - a period of accelerated expansion that occurred near the Planck time~\cite{Fomin:2018, Chervon:2019sey, Baumann:2014nda}.

Thus, modern cosmological models imply the existence of an inflationary stage of accelerated expansion in the early universe. A scalar field plays a fundamental role in the simplest models of cosmological inflation based on the Einstein gravity theory. It drives the accelerated expansion of the early universe and serves as the origin of both radiation and matter. In addition, the scalar field, cosmological perturbations source, is responsible for the formation of large-scale structure as well as the anisotropy and polarization observed in the cosmic microwave background (CMB)~\cite{Fomin:2018, Chervon:2019sey, Baumann:2014nda}.

In this paper we use the following system of units: $8\pi G = m_{p}^{-2} = c = 1$. Cosmological models of the early universe can be described by the action
\begin{align}
\label{eq:action}
S = \int d^{4}x \left( L_{g} + L_{m} + L_{int} \right),
\end{align}
where $L_{g}$ is the Lagrangian density gravitational part. It can be decomposed in terms of curvature~\cite{Baumann:2014nda}:
\begin{align}
\label{eq:Lg}
L_{g}(R) = \Lambda + \frac{1}{2}R + \alpha_{1}R^{2} + \alpha_{2}R_{\mu\nu}R^{\mu\nu} + \alpha_{3}R_{\mu\nu\rho\sigma}R^{\mu\nu\rho\sigma} + \dots,
\end{align}
where $R$ represents the scalar curvature, $R_{\mu\nu}$ denotes the Ricci tensor, $R_{\mu\nu\rho\sigma}$ --- Riemann tensor, and $\alpha_{k}$ are some constant coefficients.

The zeroth term of the decomposition $L_{g}(0) = \Lambda$ corresponds to the cosmological constant, the second term of the decomposition corresponds to Einstein gravity theory, and the other terms of this decomposition correspond to the higher curvature corrections.

Let us consider the simplest models of cosmological inflation. In this case the material part of the Lagrangian is defined as~\cite{Fomin:2018, Chervon:2019sey, Baumann:2014nda}
\begin{align}
\label{eq:Lm}
L_{m} = -\frac{1}{2}g^{\mu\nu}\partial_{\mu}\phi\partial_{\nu}\phi - V(\phi),
\end{align}
where $\phi$ is some scalar field, $V(\phi)$ is the scalar field potential, $g^{\mu\nu}$ is the metric tensor of space-time.

Non-minimal coupling between scalar field and curvature is determined by $L_{int}$. In order to reduce the model to Einstein gravity, $L_{int} = 0$.

One of the convenient methods for analyzing cosmological scenarios is measurements of the CMB anisotropy and polarization. They are crucial in the context of verifying cosmological models. As a basis, PLANCK satellite's and another experiments' data are used. They impose constraints on the spectral parameters of cosmological perturbations.

Measurements of PLANCK satellite and other datasets allow to impose constraints on the spectral cosmological perturbations' parameters. Current limits are \cite{Planck:2018, Galloni:2023}
\begin{align}
\label{eq:PS}
\mathcal{P}_{S} &= 2.1 \times 10^{-9}, \\
\label{eq:nS}
n_{S} &= 0.9649 \pm 0.004, \\
\label{eq:r}
r &= \left( \frac{\mathcal{P}_{T}}{\mathcal{P}_{S}} \right) < 0.028,
\end{align}
for the power spectrum of scalar perturbations, the spectral index of the scalar perturbations and the tensor-to-scalar ratio accordingly, $\mathcal{P}_{T}$ is the power spectra of tensor perturbations.

It should be noted that modern inflationary models consider various scalar field potentials, which impose different evolutionary nature \cite{Martin:2013tda}. The approaches have been applied to update observational limits \cite{Martin:2013tda} leading to exclude certain models, yet a significant variety of scenarios continue to satisfy all empirical requirements. Thus, an actual direction of studying cosmological inflationary models is the development of methods to enable the derivation of additional relationships between the considered parameters of the model.

One of the solutions of the problem is considering a classification between cosmological parameters. In \cite{Fomin:2024}, the models were categorized by the expansion degree of the tensor-scalar ratio's dependence on the spectral index $r = r\left( 1 - n_{S} \right)$. This expansion can be decomposed:
\begin{align}
\label{eq:r_expansion}
r = \sum_{k=0}^{\infty} \beta_{k} \left( 1 - n_{S} \right)^{k} = \beta_{0} + \beta_{1}\left( 1 - n_{S} \right) + \beta_{2}\left( 1 - n_{S} \right)^{2} + \dots,
\end{align}
where $\beta_{k}$ are the constant coefficients. By applying this approach, when a cosmological model satisfies observational constraints at a given expansion order, higher-order terms may be neglected as their contribution to the tensor-scalar ratio becomes insignificant.

Also, in \cite{Fomin:2024} a linear dependence $r \sim \left( 1 - n_{S} \right)$ was investigated for the Einstein gravity case. Such inflationary models do not correspond to observational constraints. Notice that at least the second order term $r \sim \left( 1 - n_{S} \right)^{2}$ of expansion for models is verified by observational data.

Inflationary models with a quadratic relationship between the parameters of cosmological perturbations were considered earlier in \cite{Kallosh:2013, Mishra:2018, Fomin:2019} and \cite{Odintsov:2023}. Another approach to solving the problem of determining models verified from observational data was proposed in \cite{Sousa:2024}. The study addresses parameter optimization in inflationary models through machine learning approaches applied to approximated cosmological dynamics equations, enhancing consistency with observational constraints. As a result, various types of optimal scalar field potentials for verifiable inflationary models were obtained \cite{Sousa:2024}.

In this paper we develop a reconstruction method for determining the scalar field potential type from its initial e-fold dependence, incorporating the quadratic relation between cosmological perturbation parameters. Such dependence is derived from exact solutions of cosmological dynamics equations. The effectiveness of optimization method is evaluated by obtained results for assessing inflationary model parameters against observational data within the slow-roll approximation framework.

\section{Materials and Methods}
Inflationary scenarios in general relativity are governed by an action~\cite{Fomin:2018, Chervon:2019sey, Baumann:2014nda}
\begin{align}
\label{eq:action_gr}
S = \int d^{4}x \sqrt{-g} \left[ \frac{1}{2}R - \frac{1}{2}g^{\mu\nu}\partial_{\mu}\phi\partial_{\nu}\phi - V(\phi) \right].
\end{align}
Here, we adopt unit system $8\pi G = m_{p}^{-2} = c = 1$. In Eq.~(\ref{eq:action_gr}) $\phi$ is the scalar field, $V(\phi)$ is the scalar field potential, and $g^{\mu\nu}$ is the metric tensor of the space-time.

The spatially-flat metric in the form
\begin{align}
\label{eq:metric}
ds^{2} = - dt^{2} + a^{2}(t)\delta_{ij}dx^{i}dx^{j},
\end{align}
where $a = a(t)$ is the scale factor, and $t$ is cosmic time, describes the homogenous and isotropic universe (Friedman-Robertson-Walker metric).

The dynamics equations corresponding to action can be written in the following form~\cite{Fomin:2018, Chervon:2019sey, Baumann:2014nda}
\begin{align}
\label{eq:V_H}
V\left( \phi(t) \right) &= 3H^{2} + \dot{H}, \\
\label{eq:phi_dot}
{\dot{\phi}}^{2} &= - 2\dot{H},
\end{align}
where the dot above indicates the derivative on cosmic time, and $H = \frac{\dot{a}}{a}$ is the Hubble parameter.

Also, we note, the Hubble parameter derivative
\begin{align}
\label{eq:Hdot}
\dot{H} = \frac{dH(\phi)}{d\phi} = H_{\phi}'\dot{\phi},
\end{align}
allows to represent equations (\ref{eq:V_H})-(\ref{eq:phi_dot}) in terms of the scalar field
\begin{align}
\label{eq:V_phi}
V(\phi) &= 3H^{2} - 2H_{\phi}^{'2}, \\
\label{eq:phi_dot_2}
\dot{\phi} &= 2H_{\phi}'.
\end{align}

Further, let us consider the e-folds number
\begin{align}
\label{eq:N}
N = \int_{t_{i}}^{t_{e}} H dt,
\end{align}
where $t_{i}$ and $t_{e}$ is the beginning and end time of the inflationary stage.

Thus, the derivative by the cosmic time can be rewritten in terms of the e-fold number:
\begin{align}
\label{eq:N_deriv}
\dot{N} = H, \quad \frac{d}{dt} = H \frac{d}{dN}.
\end{align}

In order to analyze inflationary dynamics, the slow-roll parameters are used. Also, they allow to evaluate primordial perturbation spectra~\cite{Fomin:2018, Chervon:2019sey, Baumann:2014nda}.

The first parameter $\varepsilon$ in slow-roll approximation is defined as follows
\begin{align}
\label{eq:epsilon_def}
\varepsilon = - \frac{\dot{H}}{H^{2}} = - \frac{1}{H} \left( \frac{dH}{dN} \right).
\end{align}
Thus, the Hubble parameter as the function of the e-folds number can be written as
\begin{align}
\label{eq:H_N}
H(N) = \lambda^{2} \exp\left( - \int \varepsilon(N) dN \right),
\end{align}
where $\lambda^{2} > 0$ is the integration constant corresponding to condition $H > 0$.

Further, we write down the equations governing cosmological dynamics in the following form
\begin{align}
\label{eq:H_phiN}
H(N) &= \lambda^{2} \exp\left( - \frac{1}{2} \int \left[ \frac{d\phi(N)}{dN} \right]^{2} dN \right), \\
\label{eq:V_phiN}
V\left( \phi(N) \right) &= \lambda^{4} \left\{ 3 - \frac{1}{2} \left[ \frac{d\phi(N)}{dN} \right]^{2} \right\} \exp\left( - \int \left[ \frac{d\phi(N)}{dN} \right]^{2} dN \right), \\
\label{eq:epsilon_N}
\varepsilon(N) &= \frac{1}{2} \left[ \frac{d\phi(N)}{dN} \right]^{2}.
\end{align}

The second slow-roll parameter can be determined as~\cite{Fomin:2018, Chervon:2019sey, Baumann:2014nda}
\begin{align}
\label{eq:delta_def}
\delta = - \frac{\ddot{H}}{2\dot{H}H} = \varepsilon - \frac{\dot{\varepsilon}}{2H\varepsilon}.
\end{align}
Based on expression (\ref{eq:epsilon_def}) it can be rewritten in the following form
\begin{align}
\label{eq:delta_N}
\delta = \varepsilon - \frac{1}{2\varepsilon} \left( \frac{d\varepsilon}{dN} \right).
\end{align}

We can also rewrite the second slow-roll parameter as
\begin{align}
\label{eq:delta_phiN}
\delta = - \left[ \frac{d\phi(N)}{dN} \right]^{2} - \frac{d}{dN} \ln\left( \frac{d\phi(N)}{dN} \right),
\end{align}
on the basis of expression (\ref{eq:epsilon_N}).

Also, when the slow-roll conditions $\varepsilon \ll 1$, $|\delta| \ll 1$ are met, the parameters of cosmological perturbations can be defined as follows ~\cite{Fomin:2018, Chervon:2019sey, Baumann:2014nda}
\begin{align}
\label{eq:PS_def}
\mathcal{P}_{S} &= \frac{1}{2\varepsilon_{*}} \left( \frac{H_{*}}{2\pi} \right)^{2}, \\
\label{eq:PT_def}
\mathcal{P}_{T} &= 8 \left( \frac{H_{*}}{2\pi} \right)^{2}, \\
\label{eq:nS_def}
1 - n_{S} &= 4\varepsilon_{*} - 2\delta_{*}, \\
\label{eq:r_def}
r &= \frac{\mathcal{P}_{T}}{\mathcal{P}_{S}} = 16\varepsilon_{*},
\end{align}
where $H_{*}$, $\varepsilon_{*}$, $\delta_{*}$ are the values of the Hubble parameter and slow-roll parameters at the crossing of the Hubble radius $(k = aH)$.

Consequently, the validation procedure for inflationary models against observational data involves comparing spectral parameters of cosmological perturbations with their observational constrained values (\ref{eq:PS})-(\ref{eq:r}).

When the slow-roll conditions $\varepsilon \ll 1$, $|\delta| \ll 1$ are met, parameters of the cosmological perturbations are defined in terms of the slow-roll parameters as
\begin{align}
\label{eq:r_epsilon}
r &= 16\varepsilon = 8 \left[ \frac{d\phi(N)}{dN} \right]^{2}, \\
\label{eq:1-nS}
1 - n_{S} &= 4\varepsilon - 2\delta = 2\varepsilon - \frac{1}{\varepsilon} \left( \frac{d\varepsilon}{dN} \right) \nonumber \\
&= \left[ \frac{d\phi(N)}{dN} \right]^{2} + 2 \left[ \frac{\frac{d^{2}\phi(N)}{dN^{2}}}{\frac{d\phi(N)}{dN}} \right] = \frac{r}{8} + 2 \left[ \frac{\frac{d^{2}\phi(N)}{dN^{2}}}{\frac{d\phi(N)}{dN}} \right],
\end{align}
where $\left( N = N_{*} \right)$ indicates crossing of the Hubble radius. In \cite{Fomin:2024, Kallosh:2013} the linear dependence $r \sim \left( 1 - n_{S} \right)$ is considered. It was shown that such inflationary models do not satisfy observational constraints. For the case of the Einstein gravity theory, the condition of verifiability can be formulated as follows
\begin{align}
\label{eq:verifiability}
\left[ \left( \frac{d^{2}\phi(N)}{dN^{2}} \right) \left( \frac{d\phi(N)}{dN} \right)^{-1} \right]_{N = N_{*}} \gg \frac{1}{2} \left. \left[ \frac{d\phi(N)}{dN} \right]^{2} \right|_{N = N_{*}} = \varepsilon\left( N = N_{*} \right) = \varepsilon_{*},
\end{align}
where Eq.~(\ref{eq:verifiability}) corresponds to the condition $\varepsilon_{*} \ll \left| \delta_{*} \right|$.

Hence, given a defined scalar field dependence relation on the number of e-folds $\phi = \phi(N)$, based on expressions (\ref{eq:H_phiN})-(\ref{eq:epsilon_N}), the type of the potential can be reconstructed corresponding to observational constraints on the parameters of cosmological perturbations.

It should also be noted that under the slow-roll condition $\varepsilon \ll 1$ the scalar field potential can be written as
\begin{align}
\label{eq:V_approx}
V\left( \phi(N) \right) = H^{2}(3 - \varepsilon) \approx 3H^{2}(N) = 3\lambda^{4} \exp\left( - \int \left[ \frac{d\phi(N)}{dN} \right]^{2} dN \right).
\end{align}
This expression makes it possible to compare the type of potential for a certain dependence $\phi = \phi(N)$ with the optimal potentials obtained in \cite{Sousa:2024}.

\section{Results}
As an example of the proposed approach for constructing models of cosmological inflation verified by the observational data, let us consider a model with a logarithmic dependence of the scalar field on the e-folds number
\begin{align}
\label{eq:phi_log}
\phi(N) = \phi_{0} + A \ln(N + B),
\end{align}
where $\phi_{0}$, $A$ and $B$ are some constants.

From expressions (\ref{eq:H_phiN})-(\ref{eq:V_phiN}) we obtain
\begin{align}
\label{eq:H_log}
H(N) &= \lambda^{2} \exp\left[ \frac{A^{2}}{2(N + B)} \right], \\
\label{eq:V_log}
V(\phi) &= 3\lambda^{4} \left( 1 - \frac{A^{2}}{6} e^{- \frac{2\left( \phi - \phi_{0} \right)}{A}} \right) \exp\left( A^{2} e^{- \frac{\left( \phi - \phi_{0} \right)}{A}} \right).
\end{align}

Further, from the expression (\ref{eq:H_log}) we get
\begin{align}
\label{eq:dHdN}
\frac{dH}{dN} = \frac{\dot{H}}{H} = - \frac{\lambda^{2} A^{2}}{2(N + B)^{2}} \exp\left[ \frac{A^{2}}{2(N + B)} \right] < 0, \quad \dot{H} < 0,
\end{align}
which corresponds to the condition $\varepsilon > 0$.

Also, from equations (\ref{eq:epsilon_N}) and (\ref{eq:delta_phiN}) one has
\begin{align}
\label{eq:epsilon_log}
\varepsilon &= \frac{A^{2}}{2(N + B)^{2}} \ll 1, \\
\label{eq:delta_log}
\delta &= \frac{A^{2}}{2(N + B)^{2}} + \frac{1}{N + B} \approx \frac{1}{N + B} \ll 1,
\end{align}
for $(N + B) \gg 1$ and $|A| \sim 1$.

We note that the slow-roll conditions are satisfied at the inflationary stage before the cosmological perturbations cross the Hubble radius. Then, the slow-roll conditions are violated, which corresponds to the exit from inflation~\cite{Fomin:2018, Chervon:2019sey, Baumann:2014nda}.

Then, when slow-roll parameters are~\cite{Fomin:2018, Chervon:2019sey, Baumann:2014nda}
\begin{align}
\label{eq:epsilon_e}
\varepsilon_{e} &= \frac{A^{2}}{2\left( N_{e} + B \right)^{2}} = 1, \\
\label{eq:delta_e}
\delta_{e} &= \frac{1}{N_{e} + B} = 1,
\end{align}
the inflationary stage is terminated and the accelerated expansion of the early universe ends. From these expressions we get $A = \pm \sqrt{2}$, which corresponds to condition $|A| \sim 1$ and $B = 1 - N_{e}$, where $N_{e}$ defines the e-folds number at the inflationary stage end.

Thus, the exact background parameters of the proposed cosmological model can be defined as follows
\begin{align}
\label{eq:phi_exact}
\phi(N) &= \phi_{0} \pm \sqrt{2} \ln\left( 1 + N - N_{e} \right), \\
\label{eq:H_exact}
H(N) &= \lambda^{2} \exp\left[ \left( 1 + N - N_{e} \right)^{-1} \right], \\
\label{eq:V_exact}
V(\phi) &= \lambda^{4} \left( 3 - e^{\mp \sqrt{2} \left( \phi - \phi_{0} \right)} \right) \exp\left( 2 e^{\mp \frac{\left( \phi - \phi_{0} \right)}{\sqrt{2}}} \right).
\end{align}

Also, from (\ref{eq:V_approx}) we obtain the scalar field potential in the slow-roll approximation
\begin{align}
\label{eq:V_slowroll}
V(\phi) \approx 3\lambda^{4} e^{A^{2}} \exp\left( e^{- \frac{\left( \phi - \phi_{0} \right)}{A}} \right) = 3\lambda^{4} e^{2} \exp\left( e^{\mp \frac{\left( \phi - \phi_{0} \right)}{\sqrt{2}}} \right),
\end{align}
which coincides with the optimal potential $V_{m}(\phi) \sim \exp\left( e^{- \phi} \right)$, obtained in \cite{Sousa:2024} for $A = \pm 1$.

Further, from (\ref{eq:r_epsilon}) and (\ref{eq:1-nS}) we obtain
\begin{align}
\label{eq:r_quadratic}
r = 2A^{2} \left( 1 - n_{S} \right)^{2} = 4 \left( 1 - n_{S} \right)^{2},
\end{align}
with the following tensor-to-scalar ratio
\begin{align}
\label{eq:r_range}
0.004 \leq r \leq 0.006,
\end{align}
which corresponds to observational constraints. However, it differs from the result obtained in \cite{Sousa:2024} $r_{m} = 0.002$ for $A = \pm 1$.

The difference in the above estimates is due to the fact that in \cite{Sousa:2024} the slow-roll parameters were determined approximately, while in this work the exact expressions (\ref{eq:epsilon_log}) and (\ref{eq:delta_log}) were used.

Likewise considering the difference between the values $\delta_{e}$ and $\delta_{*}$ (the Hubble radius crossing), from (\ref{eq:delta_log}) and (\ref{eq:1-nS}) we obtain
\begin{align}
\label{eq:DeltaN}
\frac{1}{\delta_{e}} - \frac{1}{\delta_{*}} \approx N_{e} - N_{*} = \Delta N = \frac{n_{S} - 3}{n_{S} - 1}.
\end{align}
Thus, taking into account constraint (\ref{eq:nS}), the e-folds number between the crossing of the Hubble radius by cosmological perturbations and the completion of inflation is $52 \leq \Delta N \leq 66$, which corresponds to current observational constraints \cite{Planck:2018}.

If $52 \leq \Delta N \leq 66$, the values of the slow-roll parameters at the crossing of the Hubble radius are
\begin{align}
\label{eq:epsilon_star}
\varepsilon_{*} &= \left( 1 + N_{*} - N_{e} \right)^{-2} = (1 - \Delta N)^{-2}, \quad 0.0002 \leq \varepsilon_{*} \leq 0.0004, \\
\label{eq:delta_star}
\delta_{*} &= \left( 1 + N_{*} - N_{e} \right)^{-1} = (1 - \Delta N)^{-1}, \quad -0.019 \leq \delta_{*} \leq -0.015,
\end{align}
which corresponds to condition $\varepsilon_{*} \ll \left| \delta_{*} \right|$ and the slow-roll conditions $\varepsilon \ll 1$, $|\delta| \ll 1$ as well.

Thus, the proposed model is in good agreement with observational data and can be considered as a relevant one in describing the stage of cosmological inflation.

\section{Discussion}
In this paper, a model of cosmological inflation with a quadratic relationship between the parameters of cosmological perturbations $r \sim \left( 1 - n_{S} \right)^{2}$ is proposed that corresponds to the observational data, which differs from the previously proposed models with this type of dependence $r = r\left( 1 - n_{S} \right)$. Nevertheless, the proposed type of scalar field potential in the slow-roll approximation corresponds to the optimal potential, which was obtained in \cite{Sousa:2024}. As a result, estimates of the gravitational waves contribution to the anisotropy and polarization of the CMB \cite{Baumann:2014nda, Planck:2018, Galloni:2023, Osetrin:2024} differ by a factor $\frac{r}{r_{m}} = 2$ from the model proposed in \cite{Sousa:2024}. This discrepancy arises because, unlike reference \cite{Sousa:2024}, our analysis incorporates exact solutions to the cosmological dynamics equations and precise formulations of the slow-roll parameters.

\section{Conclusion}
The considered scalar field potential reconstruction framework, based on the dependence of a scalar field on e-folds number, allows to generate verifiable inflationary models through exact solutions of cosmological dynamics equations. Our proposed approach yields inflationary model parameters that differ from those obtained via traditional slow-roll optimization methods. By comparing these results with reference \cite{Sousa:2024}, we establish their values as selection criteria for model background parameters. Nevertheless, discrepancies in the estimates of cosmological perturbation parameters correspond to the need to analyze exact cosmological solutions when verifying inflationary models by observational data.

Also, as part of the analysis of cosmological inflationary models with a scalar field \cite{Zhuravlev:2021}, special consideration should be given to different modifications of Einstein gravity theory \cite{Chirkov:2021a, Chirkov:2021b, Chervon:2023}, as these may introduce significant corrections to both the background parameters of cosmological models and the predicted values of cosmological perturbation parameters.

\section*{Acknowledgements}
I. V. Fomin's work was carried out within the framework of the Agreement on the provision of a subsidy from the federal budget for financial support for the implementation of a state assignment for the provision of public services (performance of work) No. 073-03-2025-066 dated 01/16/2025, concluded between the Federal State Budgetary Educational Institution of Higher Education "UlSPU I.N. Ulyanov" and the Ministry of Education of the Russian Federation.

\bibliography{references}

\end{document}